\documentclass[12pt]{article}
\usepackage{amsmath,amsthm,amsfonts,amscd,eucal,latexsym,amssymb} 

\newtheorem{Definition}{Definition}
\newtheorem{Theorem}[Definition]{Theorem}
\newtheorem{Proposition}[Definition]{Proposition}

\newtheorem{Corollary}[Definition]{Corollary}

\newcommand{\W}{\mathcal{W}}
\newcommand*{\hil}{\mathcal{H}}
\DeclareMathOperator{\re}{Re}
\DeclareMathOperator{\im}{Im}
\DeclareMathOperator{\dom}{dom}

\begin{document}
\begin{center}
{\Large\bf Continuity of KMS States\\[8pt]
 for Quantum Fields on Manifolds}
\\[40pt]
{\large\sc Jacek Damek}\\[30pt]
Institute of Physics,\\
University of Zielona G\'ora,\\
ul. Szafrana 4a,\\
65--516 Zielona G\'ora, Poland\\[4pt]
e-mail: J.Damek@if.uz.zgora.pl
\end{center}
${}$\\[35pt]
\noindent
{\small{\bf Abstract.}
We show that pure, quasifree states, as well as regular (i.e., those with a unique vacuum) quasifree ground and KMS states, for  linear quantum fields in a curved spacetime, are always continuous in the sense of distributions, and provide certain applications of this fact.
}
${}$\\[15pt]
Vast majority of papers on quantum fields in a curved spacetime
 exploit an assumption that  the two-point functions of the fields are distributions (see, e.g., \cite{passive,Schlieder,stroh,analitic}). This property is, for instance, crucial for powerful methods of microlocal analysis and, consequently, for the study of Hadamard states. Nevertheless, this fact does not seem to have been established generally for the most prominent states of the physical theory; namely, ground and KMS states. It is even thought somewhere in the literature that all quasifree states are regular in the sense concerned. However, that is  not the case, as may be seen from  examples in this article. 

 In the setting of a model linear scalar field theory, we prove the continuity for quasifree pure states and, subsequently, demonstrate this feature to hold for quasifree regular ground and KMS states, thereby exposing its inherent character, independent of any peculiarities of operator theory usually involved here. These results allow one, for instance, to drop the continuity assumption, in the above cases, in a remarkable theorem due to Sahlmann and Verch \cite{passive} and claim that regular ground and KMS states of a scalar field are always of the Hadamard form. We also provide certain exemplary consequences of the established continuity and point out some other useful omissions of this assumption in the literature.

We will be concerned only with a linear hermitean scalar field as possible generalizations of our results are rather straightforward. Let $(M,g)$ be a globally hyperbolic spacetime (second countable, paracompact, orientable, n--dimensional) with a time orientation chosen. Assume that $\varSigma$ is a smooth Cauchy surface in $M$. For the wave equation 
\begin{equation}
\label{wave}
\left(\square_g + V\right)\varphi = 0,\qquad V\in C^\infty(M),
\end{equation}
we may define the space, $S$, of compact Cauchy data $C^\infty_0(\varSigma) \oplus 
C^\infty_0(\varSigma)$ for solutions of (\ref{wave}), and the usual classical symplectic form $\sigma$ on $\varSigma$. Now, let $\mathcal{W}$ be the Weyl algebra of canonical commutation relations over the symplectic space $(S,\sigma)$. For any real scalar product $\mu$ on $\varSigma$, one can define a state $\omega_\mu$ on $\mathcal{W}$ \cite{K-W} by  
 \begin{equation}
\label{qfree}
\omega_\mu\left(W(\phi)\right)= \exp(-\frac{1}{2}\mu(\phi,\phi)), \qquad \phi \in S,
\end{equation}
 where $W(\phi)$ stands for  Weyl generators of $\mathcal{W}$, provided the condition
 \begin{equation}
\left(\sigma(\phi_1,\phi_2)\right)^2 \leq 4\mu(\phi_1,\phi_1)\,\mu(\phi_2,\phi_2),
\qquad \phi_1, \phi_2 \in S. \label{cond}
 \end{equation}
A quasifree state is just a state obtained in this way. 

As is well--known \cite{K-W}, there exists a one--particle Hilbert space structure for a state $\omega_\mu$. This consists of a Hilbert space $\mathcal{H}$, and a real--linear mapping $K\colon S
\to \hil$ such that, for all $\phi_1,\phi_2 \in S$, 
\begin{enumerate}
\item[(i)]
$\mu(\phi_1,\phi_2) = \re \langle K\phi_1,K\phi_2 \rangle$,
\item[(ii)]
$\sigma(\phi_1,\phi_2) = 2 \im \langle K\phi_1,K\phi_2 \rangle$,
\item[(iii)]
$KS + iKS\;\;$is dense in $\hil$.
\end{enumerate}
In the sequel, we shall also need the fact that $KS$ is dense in $\hil$ itself if and only if a state $\omega_\mu$ is pure \cite{K-W}.

After those preliminaries, we are now able to present our definition of continuity of states on
$\W$, as well as its convenient equivalents in the forthcoming proposition. 
\begin{Definition}
A state $\omega$ over $\mathcal{W}$ will be called continuous if he function $\phi \to \omega
\left(W(\phi)\right)$ is continuous w.r.t. the Schwartz topology of $S$.
\end{Definition}
In the quasifree case, this definition amounts to the so--called $C^\infty$--regularity of a state (cf. \cite{vector}), which, notably, ensures the existence of Wightman disributions for 
the quantum field theory resulting from such a state via the GNS construction.
\begin{Proposition}
\label{equiv}
For a quasifree state $\omega_\mu$, the following conditions are equivalent:
\begin{enumerate}
\item[(i)]
$\omega_\mu$ is continuous,
\item[(ii)]
$\mu$ is continuous on $S\times S$,
\item[(iii)]
$K$ is continuous (in  the norm of $\hil$).
\end{enumerate}
\end{Proposition} 
This easily follows from the formula (\ref{qfree}), polarization for $\mu$, and condition (i)
of the definition of a one--particle structure for $\mu$. 

 The inequality (\ref{cond}) forces the continuity of $\sigma$ whenever a state associated with $\mu$ is continuous, but the converse is false. Take an arbitrary $\mu_0$ defining a continuous quasifree state and, therefore, satisfying (\ref{cond}), and some discontinuous linear functional $s$ on $S$. Then $\mu_s = \mu_0 + s \otimes s$ defines a discontinuous quasifree state on $\W$, since such a $\mu_s$ clearly fulfills the desired condition (\ref{cond}) if $\mu_0$ does. In view of the next theorem, we note that this exemplary singular state is not a pure state.

We now state and prove our basic theorem.
\begin{Theorem} \label{pure}
The pure quasifree states on $\W$ are continuous.
\end{Theorem}
\begin{proof}
Let $\omega_\mu$ be an arbitrary pure quasifree state with a one--particle Hilbert space structure $(\hil,K)$. By Proposition (\ref{equiv}), it suffices to prove the  continuity for the mapping $K$. For any compact set $C \subset \Sigma$, we define $S_C \subset S$ by
\[ S_C = C^\infty_0(C) \oplus C^\infty_0(C).\]
Then $S_C$ is a Frechet (metrizable) space. We will prove that $K \upharpoonright S_C$ is continuous, which, of course, implies the desired continuity of $K$ on all of $S$.
To show this, we employ the closed graph theorem for the spaces $S_C$
and $\hil$. 

Accordingly, suppose that $\phi_n \to \phi,\ K\phi_n \to x\ (\phi_n \in S_C,\; x \in \hil)$ with obvious meaning. We only need to show that $x = K\phi$. By the definition of $K$, for any 
$\psi \in S$, we have 
\begin{equation}
\label{im}
\sigma(\psi,\phi_n) = 2 \im \langle K\psi,K\phi_n \rangle.
\end{equation}
Passing to the limit yields 
 \begin{equation*}
\sigma(\psi,\phi) = 2 \im \langle K\psi,x \rangle,
\end{equation*}
as the classical form $\sigma$ is continuous.
On applying (\ref{im}) with $\phi$ instead of $\phi_n$, we then obtain
\begin{equation}
\label{im2}
 \im \langle K\psi,K\phi \rangle =  \im \langle K\psi,x \rangle.
\end{equation}
By purity of the state, there is a sequence $\{\psi_n\} \in S$ such that $iK\psi = \lim\limits_{n \to \infty} K\psi_n$. Substituting $K\psi_n$ for $K\psi$ in (\ref{im2}) and taking the limit, we arrive at
\begin{equation} 
\label{re}
 \re \langle K\psi,K\phi \rangle =  \re \langle K\psi,x \rangle.
\end{equation}
 Finally, it follows from (\ref{im2}), (\ref{re}), and the denseness of $KS$ in $\hil$ that 
$K\phi = x$, which completes the proof.
\end{proof}
Now, we shall infer a few consequences of the above general theorem, which are of more direct physical interest. To this end, we assume that our spacetime is stationary, which allows one to define the quasifree symplectic time evolution on the Weyl algebra considered here. Then, the standard notions of ground or KMS states make sense with respect to the just introduced time evolution. Further, let us explain that by regular quasifree ground or KMS  states we mean those whose one--particle hamiltonian has no zero modes. 
\begin{Corollary} 
The regular, quasifree ground states are continuous.
\end{Corollary}
This is true due to Theorem \ref{pure} and the fact that such states are pure \cite{K-W}.
\begin{Proposition}
The regular, quasifree KMS states are continuous.
\end{Proposition}
\begin{proof}
Such a KMS state, say $\omega$, necessarily arises from a unique regular ground one--particle Hilbert space structure $(\hil, K, h)$, where h is a one--particle hamiltonian (see \cite{K-W,Kay}). Next, we may double  the underlying spacetime $M$, so as to obtain a stationary spacetime $\tilde{M} = M_L \cup M_R\:$,\ $M_L,\: M_R$ being copies of $M$. Then, on the Weyl algebra for $\tilde{M}$, one can build a double quasifree KMS state by doubling the structure $(\hil, K, h)$ (see \cite{Kay}) in order to obtain the double KMS one--particle structure in $\hil \oplus \hil$, which then defines that state. This state is both quasifree and pure, and coincides with $\omega$ on $M_R$. Thus, by Theorem \ref{pure}, the state $\omega$ is continuous.
\end{proof}
In view of, e.g., Lemma 6.1 in ref. \cite{K-W}, we note that for a one--particle Hilbert space structure $(\hil, K, h)$ of such a KMS or ground state, the condition $KS \subset \dom (h)$  always holds. To see this, we write 
\begin{equation*}
\frac{d}{dt}\, e^{iht}K\phi = \frac{d}{dt} K\phi_t = K\frac{d}{dt} \phi_t,
 \end{equation*}
where $\phi_t$ stands for the time translate of $\phi$. The existence of the derivative inside the formula is guaranteed by the continuity of $K$ and the fact that $\frac{\phi_t - \phi}{t}$ converges to $\frac{d}{dt}\phi_t$ in the Schwartz topology of $S$. 

In the case of a continuous quasifree state, the one--particle Hilbert space of that state (thus the Fock space of the GNS representation) must be separable. This stems from continuity of $K$, denseness of $KS + iKS$, and separability  of $S_C$ (the space having been defined at the beginning of the proof of Theorem \ref{pure}) together with the fact that a Cauchy surface in our spacetime is a sum of a countable number of compact sets. 

	Although there is no doubt that explicitly constructed states, such as those for a Schwarzschild black hole rigorously introduced by Kay \cite{Kay}, are continuous, now this property has been shown to hold almost automatically, apart from any assumptions on self--adjointness or spectral expansions. This can be helpful, especially in the context of abstract, general theorems, where one can thus drop the continuity assumption in the above-noted, common cases. For example, Sahlmann and Verch's \cite{passive} general theorem on the microlocal spectrum condition for passive (e.g., ground or KMS) states implies, in view of  Radzikowski's result \cite{Radzik}, that these states, if defined on our Weyl algebra, are of the Hadamard form on condition that their two--point functions are continuous. However, we may drop this requirement in the important case of regular, quasifree states, as explained in this article. 

	In ref. \cite{Schlieder} the Reeh--Schlieder property has been established for quasifree, ground and KMS states under the continuity assumption, which may again be dropped for regular states.

Finally, we remark that there is no obstacle to generalize our results to vector--valued fields of Verch's type \cite{vector}.\\[3ex]

It is pleasure to thank Roman Gielerak for discussions related to this work.

\end{document}